\documentclass[a4paper]{article}
\usepackage[latin1]{inputenc}
\sf
\usepackage{amssymb}

\def\C{\mathbb{C}}

\usepackage{amssymb}

\def\C{\mathbb{C}}

\newtheorem{theo}{Theorem}
\newtheorem{defi}{Definition}

\begin{document}
\title{Explicit Formulae for Cocycles of Holomorphic Vector Fields
with values in $\lambda$ Densities}
     
\author{
        Friedrich Wagemann \\
        Institut Girard Desargues -- UPRES-A 5028 du CNRS\\
	Universit\'e Claude Bernard Lyon-I\\
        43, bd du 11. Novembre 1918 \\
	69622 Villeurbanne Cedex FRANCE\\
        tel.: +33.4.72.43.11.90\\
        fax:  +33.4.72.43.00.35\\
        e-mail: wagemann@desargues.univ-lyon1.fr}
\maketitle


\begin{abstract}
In this article, we give explicit formulae for the generators of
$H^2(Hol(\Sigma),{\cal F}_{\lambda}(\Sigma))$ in terms of affine and
projective connections. This is done using the cocycles which have been
evidenced by V. Ovsienko and C. Roger in \cite{OvsRog} and globalizing
them by their transformation property.
\end{abstract} 
 
\section*{Introduction}

The continuous cohomology of Lie algebras of ${\cal
C}^{\infty}$-vector fields has been studied by I. M. Gelfand,
D. B. Fuks, R. Bott, A. Haefliger and G. Segal in some outstanding
papers \cite{GelFuk}, \cite{Hae}, \cite{BotSeg}.

B. L. Feigin \cite{Fei} and N. Kawazumi \cite{Kaw}, whose work is
continued in \cite{Wag1}, studied
Gelfand-Fuks cohomology of Lie algebras of holomorphic vector fields
$Hol(\Sigma)$ on an open Riemann surface. Kawazumi calculated the
cohomology spaces $H^*(Hol(\Sigma),{\cal F}_{\lambda}(\Sigma))$ of
$Hol(\Sigma)$ with values in the space of (holomorphic) $\lambda$-densities on
$\Sigma$, using a well known theorem of Goncharova, cf
\cite{Fuk}. He expressed the generators of the
cohomology spaces in terms of the nowhere-vanishing holomorphic vector
field $\partial$ which exists on open Riemann surfaces, trivializing
the holomorphic tangent bundle.

In this article, we give explicit formulae for the generators of
$H^2(Hol(\Sigma),{\cal F}_{\lambda}(\Sigma))$ in terms of affine and
projective connections. This is done using the cocycles which have been
evidenced by V. Ovsienko and C. Roger in \cite{OvsRog} and globalizing
them by their transformation property.

The main reason to look for explicit formulae is the search for a
generalization of the Krichever-Novikov algebras \cite{KriNov} to   
semi-direct products of $Hol(\Sigma)$ with ${\cal
F}_{\lambda}(\Sigma)$, cf \cite{OvsRog} for the case of $Vect(S^1)$.\\

{\bf Acknoledgements:}

The author thanks V. Ovsienko and C. Roger for the statement of the
problem and usefull discussions on the subject of this paper. He
thanks also D. Millionshshikov for illuminating conversations and his
formula for the Krichever-Novikov cocycle. 

\section{Preliminairies, statement of the result}

In this section, we state the theorems of Kawazumi and of
Ovsienko-Roger which are the starting point of our work.

Let $Vect(S^1)$ denote the Lie algebra of differentiable vector fields
on the circle $S^1$, and ${\cal F}_{\lambda}$ the $Vect(S^1)$-module
of $\lambda$-densities, using the action
 
\begin{equation} \label{*}
L_fa = (fa' + \lambda f'a)(dx)^{\lambda},
\end{equation}

where $f\in Vect(S^1)$ and $a\in{\cal F}_{\lambda}$ are both
represented by their coefficient function.

\begin{theo}[Theorem 3, \cite{OvsRog}]
The cohomology groups $H^2(Vect(S^1),{\cal F}_{\lambda})$ are non-zero
only for $\lambda=0,1,2,5,7$. They are two-dimensional for
$\lambda=0,1,2$ and one-dimensional for $\lambda=5,7$.

The generators read explicitly

\begin{eqnarray*}
\bar{c}_0(f,g)&=&\left|\begin{array}{cc}f & g \\ f' &
g'\end{array}\right| \\
c_0(f,g)&=&c_{GF}(f,g) \\
c_1(f,g)&=&\left|\begin{array}{cc}f' & g' \\ f'' &
g''\end{array}\right|dx \\
\bar{c}_1(f,g)&=&\left|\begin{array}{cc}f & g \\ f'' &
g''\end{array}\right|dx \\
c_2(f,g)&=&\left|\begin{array}{cc}f' & g' \\ f''' &
g'''\end{array}\right|(dx)^2 \\
\bar{c}_2(f,g)&=&\left|\begin{array}{cc}f & g \\ f''' &
g'''\end{array}\right|(dx)^2 \\
c_5(f,g)&=&\left|\begin{array}{cc}f''' & g''' \\ f^{(IV)} &
g^{(IV)}\end{array}\right|(dx)^5 \\
c_7(f,g)&=&\left(2\left|\begin{array}{cc}f''' & g''' \\ f^{(VI)} &
g^{(VI)}\end{array}\right|  - 9\left|\begin{array}{cc}f^{(IV)} &
g^{(IV)} \\ f^{(V)} & g^{(V)}\end{array}\right|\right)(dx)^7 
\end{eqnarray*}

Here, $c_{GF}$ is the Gelfand-Fuks cocycle, cf \cite{Fuk}, being a
cocycle with values in the trivial module $\C\subset{\cal F}_0$.
\end{theo}

Now, let $\Sigma_r$ - as in the rest of this article - denote an open
Riemann surface, obtained from the compact Riemann surface $\Sigma$ by
extraction of $r$ points:
$\Sigma_r:=\Sigma\setminus\{p_1,\ldots,p_r\}$.

Let $Hol(\Sigma_r)$ denote the (infinite dimensional) Lie
algebra of holomorphic vector fields on $\Sigma_r$. Let ${\cal
F}_{\lambda}(\Sigma_r)$ be the space of sections of the bundle
of holomorphic $\lambda$-densities. As all bundles on $\Sigma_r$ are
trivial, elements of ${\cal F}_{\lambda}(\Sigma_r)$ can be represented
by holomorphic functions. $Hol(\Sigma)$ still acts on ${\cal
F}_{\lambda}(\Sigma_r)$ according to 

\begin{displaymath}
L_fa = (fa' + \lambda f'a)(dz)^{\lambda},
\end{displaymath}

where $f\in Hol(\Sigma_r)$ and $a\in{\cal F}_{\lambda}(\Sigma_r)$ are both
represented by their coefficient function, the $(dz)^{\lambda}$ being
the global section trivialising the bundle of $\lambda$-densities.

Recall that 

\begin{displaymath}
H^p(\Sigma_r)=\left\{\begin{array}{ccc}\C&{\rm for}&p=0 \\
\C^{2g+r-1}&{\rm for}&p=1 \\ 0&{\rm for}&p\geq 2
\end{array}\right.,
\end{displaymath}

if $g$ denotes the genus of $\Sigma$.
 
In \cite{Kaw}, Kawazumi calculates the spaces $H^2(Hol(\Sigma_r),{\cal
F}_{\lambda}(\Sigma_r))$ using the Re\u{s}etnikov \cite{Res} spectral
sequence. This sequence has as $E_2$-term the sheaf cohomology of
 a sheaf whose stalk at $x\in\Sigma_r$ is the cohomology of
$Hol(\Sigma_r)$ with values in ${\cal F}_{\lambda}(\Sigma_r)_x$, the fibre
of ${\cal F}_{\lambda}(\Sigma_r)$ at $x$.

Furthermore, Kawazumi uses the main result of his article to express
the stated $E_2$-term as a tensor product of some ``covariant
derivative'' cocycles with the formal version of his cohomology,
namely $H^*(W_1,T_{\lambda})$. Here, $W_1$ is the Lie algebra of
formal vector fields on the complex line and $T_{\lambda}$ is the
corresponding module of formal
$\lambda$-densities. $H^*(W_1,T_{\lambda})$ is explicitly given,
thanks to the theorem of Goncharova, cf \cite{Fuk}. Thus, he obtains
a (collapsing) spectral sequence for the wanted cohomology.

Let us state his result just for the dimensions of $H^2(Hol(\Sigma_r),{\cal
F}_{\lambda}(\Sigma_r))$ for the different $\lambda$. 

\begin{theo}[consequence of (9.7), \cite{Kaw}]
\begin{eqnarray*}
{\rm dim}\,\,H^2(Hol(\Sigma_r),{\cal F}_0(\Sigma_r)) &=& 2(2g+r-1)\\
{\rm dim}\,\,H^2(Hol(\Sigma_r),{\cal F}_1(\Sigma_r)) &=& 2g+r\\
{\rm dim}\,\,H^2(Hol(\Sigma_r),{\cal F}_2(\Sigma_r)) &=& 2g+r\\
{\rm dim}\,\,H^2(Hol(\Sigma_r),{\cal F}_5(\Sigma_r)) &=& 1\\
{\rm dim}\,\,H^2(Hol(\Sigma_r),{\cal F}_7(\Sigma_r)) &=& 1
\end{eqnarray*}

For all other values of $\lambda$, $H^2(Hol(\Sigma_r),{\cal
F}_{\lambda}(\Sigma_r))$ is zero.
\end{theo}

To understand these dimensions, recall from \cite{Kaw} (9.7)
p.701 that for $\lambda=0$, $H^2(Hol(\Sigma_r),{\cal
F}_{\lambda}(\Sigma_r))$ is generated by some classes,
which we denote $c^{\omega}_0$ and $\bar{c}^{\omega}_0$, each
depending on an element $\omega\in H^1(\Sigma_r)=\C^{2g+r-1}$. In the
same manner, $H^2(Hol(\Sigma_r),{\cal F}_1(\Sigma_r))$ is generated by
one family, denoted $\bar{c}_1^{\omega}$, and a cocycle $c_1$,
$H^2(Hol(\Sigma_r),{\cal F}_2(\Sigma_r))$ is generated by a family,
denoted $\bar{c}_2^{\omega}$, and a cocycle $c_2$ and
$H^2(Hol(\Sigma_r),{\cal F}_5(\Sigma_r))$ and $H^2(Hol(\Sigma_r),{\cal
F}_7(\Sigma_r))$ are each generated by a cocycle $c_5$ and $c_7$. We
have chosen the same notation as in the above theorem of Ovsienko
and Roger, but their cocycles would not give globally defined objects
on a Riemann surface $\Sigma_r$. The explicit construction in terms of
connections of these cocycles, resp. families of cocycles, is not known
and the subject of this article.

Partial results are known: namely, the holomorphic version of the
Gelfand-Fuks cocycle leading to a meromorphic version of the Virasoro
algebra appeared in work of Krichever and Novikov, further developped by
Schlichenmaier and Sheinman, cf \cite{SchShe}. It reads (cf \cite{Wag}
where we applied Poincar\'e duality to write it as an integral over
$\Sigma_r$):  

\begin{displaymath}
c_0^{\omega}(f,g) = \frac{c}{24\pi{\bf
i}}\int_{\Sigma_r}\left(\frac{1}{2}\left|\begin{array}{cc} f & g \\ f''' &
g''' \end{array}\right| - R\left|\begin{array}{cc} f & g \\ f' &
g' \end{array}\right|\right)dz\wedge\bar{\omega},
\end{displaymath}

where $\omega\in H^1(\Sigma_r)$ and $R$ is a projective
connection. Recall that for a Stein manifold (in particular for an open
Riemann surface) the subcomplex of holomorphic forms calculates all
the de Rham cohomology, cf \cite{GriHar} p.449. 

On this form, we already see how such a cocycle is constructed: the
Gelfand-Fuks cocycle serves as symbol, and then one adds terms to have
a globally defined 1-form. In other words, the 1-form without the term
involving $R$ is globally defined only with respect to an atlas of
charts from $PSl(2;\C)$, to define it for a general holomorphic atlas,
one has to use a projective connection.

Affine connections are more general than projective connections, namely,
a manifold supporting an affine connection admits also a projective
connection. These connections come from the corresponding structures, an
affine (projective) structure being an (holomorphic) atlas such that
the chart transitions are in the subgroup of affine (resp. projective)
transformations. One sees that an affine structure is in particular a
projective structure. We will state some well known facts about these
objects in the next section.

Let us state the main result of this article:

\begin{theo}
Let $\Sigma_r$ be an open Riemann surface, $Hol(\Sigma_r)$ the Lie
algebra of holomorphic vector fields on $\Sigma_r$ and ${\cal
F}_{\lambda}(\Sigma_r)$ the space of holomorphic $\lambda$-densities. 

The spaces $H^2(Hol(\Sigma_r),{\cal F}_{\lambda}(\Sigma_r))$ for
$\lambda=0,1,2,5,7$ are generated by the classes
$\bar{c}_0,c_0^{\omega}, c_1, \bar{c}_1^{\omega}, c_2,
\bar{c}_2^{\omega}, c_5, c_7$, where the subscript indicates the value
of $\lambda$ and the superscript the dependence on (the class of) a
holomorphic 1 form $\omega$ on $\Sigma_r$.

The explicit formulae are given in section 2 (in terms of affine and
projective connections) and in section 4 (in terms of the covariant
derivative).
\end{theo}

Note that the theorem asserts in particular that the formulae in
section 2 and in section 4 for $\bar{c}_0,c_0^{\omega}, c_1,
\bar{c}_1^{\omega}, c_2, \bar{c}_2^{\omega}, c_5, c_7$ coincide.

\section{Transformation behavior}

In this section, we shall calculate the correction terms in order to
make the cocycles of theorem 1 globally defined geometrical objects.

Let $X,Y$ denote holomorphic vector fields on $\Sigma_r$. Let $U_{\alpha},
U_{\beta}\subset\Sigma_r$ be open subsets such that $U_{\alpha}\cap
U_{\beta}\not=\emptyset$. Let $X$ and $Y$ be given by local
coefficient functions $f_{\alpha}, g_{\alpha}$ in $U_{\alpha}$ and
$f_{\beta}, g_{\beta}$ in $U_{\beta}$. Denote by $z_{\alpha}$ and
$z_{\beta}$ local coordinates in $U_{\alpha}$ and $U_{\beta}$, and by
$h(z_{\alpha})=z_{\beta}$ the holomorphic change of coordinates.  We have

\begin{displaymath}
f_{\beta} = \frac{\partial h}{\partial z_{\alpha}} f_{\alpha},
\end{displaymath}

and similarly for $g_{\beta}$. Denote $\frac{\partial h}{\partial
z_{\alpha}}$ just by $h'$.

Now, it is easy to transform derivatives on the coefficient functions:

\begin{displaymath}
f_{\beta}' = \frac{1}{h'}(h''f_{\alpha}) +  f_{\alpha}',
\end{displaymath}

and 
 
\begin{displaymath}
f_{\beta}'' =
\frac{1}{h'}\left\{\left(\frac{h'''}{h'}-\frac{(h'')^2}{(h')^2}\right)f_{\alpha}+\frac{h''}{h'}f_{\alpha}' + f_{\alpha}''\right\}.
\end{displaymath}

Remark that this kind of manipulations is particularly well suited for
being treated by MAPLE.

Denote by $S:=S(h)$ the Schwartzien derivative of $h$, i.e. the expression 

\begin{displaymath}
S = \frac{h'''}{h'} - \frac{3}{2}\left(\frac{h''}{h'}\right)^2.
\end{displaymath}

It is easy to show and well known that we have:

\begin{displaymath}
f_{\beta}''' = \frac{1}{(h')^2}\left(f_{\alpha}''' + S'f_{\alpha} +
2Sf_{\alpha}'\right).
\end{displaymath}

Now recall some generalities on affine and projective structures
resp. connections, cf \cite{Gun} \S 9 p. 164 and \cite{Sch}
p. 137--138:

\begin{defi}
Let $\{U_{\alpha},z_{\alpha}\}$ be a covering of $\Sigma_r$ by
coordinate charts and $z_{\beta}=h(z_{\alpha})$ the coordinate
transitions for non-empty $U_{\alpha}\cap U_{\beta}$.

A (holomorphic) projective connection is a family of holomorphic
functions $R_{\alpha}$ on $U_{\alpha}$ such that for non-empty
$U_{\alpha}\cap U_{\beta}$, we have

\begin{displaymath}
R_{\beta}(h')^2\,=\,R_{\alpha}\,+\,S.
\end{displaymath}
\end{defi}

In the same way, we have

\begin{defi}
A (holomorphic) affine connection is a family of holomorphic
functions $T_{\alpha}$ on $U_{\alpha}$ such that for non-empty
$U_{\alpha}\cap U_{\beta}$, we have

\begin{displaymath}
T_{\beta}h'\,=\,T_{\alpha}\,+\,\frac{h''}{h'}.
\end{displaymath}
\end{defi}

(Observe that $h'\not= 0$.) There is a 1-1 sorrespondance between
connections and the corresponding structures, see \cite{Gun} thm. 19,
p. 170. See also \cite{Ghy}, section 2, for a brief summary on these
structures. Affine connections (thus affine structures, projective
structures and projective connections) exist on any open Riemann
surface, cf \cite{Gun2}. This is in contrast to compact Riemann
surfaces where affine connections exist only for genus 1, see
\cite{Gun} p. 173.  

We use these objects to compensate extra terms arising from the
transition behaviour of the cocycles of theorem 1. One arrives at the
following results (the first 2 are trivial; in the following, $R$
denotes a projective connection, and $T$ an affine connection):

\begin{itemize}
\item $\bar{c}_0(f,g)$ is a well-defined global vector field, so
$\bar{c}_0^{\omega}(f,g):=\bar{c}_0(f,g)\omega$ is a well-defined
global function\\ 
\item $c_0^{\omega}(f,g)$ is a well-defined global (constant) function \\
\item $c_1(f,g):= \left|\begin{array}{cc}f' & g' \\ f'' &
g''\end{array}\right| - T\left|\begin{array}{cc}f & g \\ f'' &
g''\end{array}\right| + (R-\frac{1}{2}T^2)\left|\begin{array}{cc}f & g \\ f' &
g'\end{array}\right|$ is a well-defined global 1-form \\
\item $\bar{c}_1(f,g):= \left|\begin{array}{cc}f & g \\ f'' &
g''\end{array}\right| - T\left|\begin{array}{cc}f & g \\ f' &
g'\end{array}\right|$ is a well-defined global function, so
$\bar{c}^{\omega}_1(f,g):=\bar{c}_1(f,g)\omega$ is a well-defined
global 1-form\\ 
\item $c_2(f,g):=\left|\begin{array}{cc}f' & g' \\ f''' &
g'''\end{array}\right| - T \left|\begin{array}{cc}f & g \\ f''' &
g'''\end{array}\right| - (2TR - R')\left|\begin{array}{cc}f & g \\ f' &
g'\end{array}\right|$ is a well-defined global 2-form\\
\item $\bar{c}_2(f,g):=\left|\begin{array}{cc}f & g \\ f''' &
g'''\end{array}\right| - 2R\left|\begin{array}{cc}f & g \\ f' &
g'\end{array}\right|$  is a well-defined global 1-form, so
$\bar{c}^{\omega}_2(f,g):=\bar{c}_2(f,g)\otimes\omega$  is a
well-defined global quadratic differential \\ 
\item ${c}_5(f,g):=\left|\begin{array}{cc}f''' & g''' \\ f^{(IV)} &
g^{(IV)}\end{array}\right| + R''\left|\begin{array}{cc}f & g \\ f''' &
g'''\end{array}\right| + 3R' \left|\begin{array}{cc}f' & g' \\ f''' &
g'''\end{array}\right| + 2R\left|\begin{array}{cc}f'' & g'' \\ f''' &
g'''\end{array}\right| + (2RR'-3(R')^2)\left|\begin{array}{cc}f & g \\ f' &
g'\end{array}\right| - 2RR'\left|\begin{array}{cc}f & g \\ f'' &
g''\end{array}\right| - R'\left|\begin{array}{cc}f & g \\ f^{(IV)} &
g^{(IV)}\end{array}\right| -4R^2\left|\begin{array}{cc}f' & g' \\ f'' &
g''\end{array}\right| -2R\left|\begin{array}{cc}f' & g' \\ f^{(IV)} &
g^{(IV)}\end{array}\right|$  is a well-defined global 5-form
\end{itemize}

Note that the assignment of holomorphic 1-forms $\omega$ to certain
cocycles gives exactly the number of generators which is needed to generate the
cohomology spaces. We left out the formula for $c_7$ which is too long
to be reproduced here. 

\section{Cocycle property}

Now, we have globalized the cocycles to individual cochains or families of
cochains. But it is not clear whether the terms that we added will
disturb the cocycle property. This is what we check in this section.

By writing explicitly the cocycle identity for the different
expressions which we considered in the preceeding section to globalize
the cocycles (the action depends on $\lambda$ (cf equation
(\ref{*}) in the preliminairies)), we get the following result:

Note that the $6$th and the $7$th expressions arise as terms in
$c_5(f,g)$. Note that in this formal calculation, $f,g$ can be
interpreted as vector fields on the circle or on the open Riemann
surface. In the latter case, the expressions are not globally defined
geometric objects.  

\begin{itemize}
\item $ \left|\begin{array}{cc}f & g \\ f' & g'\end{array}\right|$ is
a cocycle for any value of $\lambda$ \\ 
\item $\left|\begin{array}{cc}f & g \\ f'' & g''\end{array}\right|$ is
a cocycle only for $\lambda=1$ \\ 
\item $\left|\begin{array}{cc}f & g \\ f''' & g'''\end{array}\right|$
is a cocycle only for $\lambda=2$ \\ 
\item $\left|\begin{array}{cc}f' & g' \\ f'' & g''\end{array}\right|$ is a cocycle only when taking trivial action\\
\item $\left|\begin{array}{cc}f' & g' \\ f''' &
g'''\end{array}\right|$ is a cocycle for any value of $\lambda$ \\ 
\item $\left|\begin{array}{cc}f & g \\ f^{(IV)} &
g^{(IV)}\end{array}\right|$ is never a cocycle \\
\item $\left|\begin{array}{cc}f' & g' \\ f^{(IV)} &
g^{(IV)}\end{array}\right|$ is never a cocycle \\
\item $\left|\begin{array}{cc}f'' & g'' \\ f''' &
g'''\end{array}\right|$ is a cocycle only for $\lambda=3$ \\
\item $\left|\begin{array}{cc}f''' & g''' \\ f^{(IV)} &
g^{(IV)}\end{array}\right|$ is a cocycle only for $\lambda=5$ 
\end{itemize}

It is thus obvious that $\bar{c}_0^{\omega}$, $c_0^{\omega}$, $c_1$,
$\bar{c}_1^{\omega}$  $c_2$,  $\bar{c}_2^{\omega}$ are well-defined,
global 2-cocycles for cohomology with values in ${\cal F}_{\lambda}$
with $\lambda=0,0,1,1,2$ and $2$ respectively.

For $c_5$ and $c_7$, we will take a different point of view.

\section{Formulation in terms of the covariant derivative}

The fundamental fact which assures the validity of our work is the
existence of affine structures on open Riemann surfaces.

These connections are flat integrable connections in the sense of
differential geometry, thus we can talk about associated covariant
derivatives. The covariant derivative associated to the affine
connection reads locally (on a $\lambda$-density $\phi$)

\begin{displaymath}
\nabla\phi\,=\,\phi'\,-\,\lambda\Gamma\phi.
\end{displaymath}

In general, $\Gamma$ plays the role of the trace of the Christoffel
symbols; in higher dimensions, we have 

\begin{displaymath}
\nabla_i\phi\,=\,\partial_i\phi\,-\,\lambda\Gamma_{ij}^j\phi.
\end{displaymath}

Actually, $\Gamma$ is nothing else than what we 
called before the affine connection $T$. $\nabla\phi$ is a globally
defined object. On $\frac{1}{2}$-densities, one can exhibit a
particular convenient choice of a projective connection associated to
an affine connection:

\begin{eqnarray*}
\nabla^2(\phi(dz)^{-\frac{1}{2}})&=&\nabla((\phi'+\frac{1}{2}\Gamma\phi)(dz)^{\frac{1}{2}})\\
 &=&(\phi''+\frac{1}{2}(\Gamma\phi)'-\frac{1}{2}\Gamma(\phi'+\frac{1}{2}\Gamma\phi))(dz)^{\frac{3}{2}}\\
 &=&(\phi''-\frac{1}{4}\Gamma^2\phi+\frac{1}{2}\Gamma'\phi)(dz)^{\frac{3}{2}}\\
& =&(\partial^2+\frac{1}{2}R)\phi(dz)^{-\frac{1}{2}}
\end{eqnarray*}

where we have put $R=-\frac{1}{2}\Gamma^2+\Gamma'$. 

Otherwise, this choice is justified by $\Gamma=\frac{h''}{h'}$ and
$S=\frac{h'''}{h'}-\frac{3}{2}\left(\frac{h''}{h'}\right)^2$, giving
also $S=\Gamma'-\frac{1}{2}\Gamma^2$, cf \cite{SchHaw} equation (10) p. 205.

Furthermore, we can set

\begin{displaymath}
L_f\phi(dx)^\lambda\,=\,f\nabla\phi\,+\,\lambda(\nabla f)\phi,
\end{displaymath}

because this action coincides with the action defined in equation (1). We
have $[f,g]=f\nabla g - (\nabla f)g$ and a derivation property of
$\nabla$ on tensor products. This corresponds to the product formula
for the derivative. With this in mind, we have the same rules of
manipulation as before for computations which concerned only
ordinary derivatives of functions on the cercle.

In conclusion, it is clear that we can formulate all cocycles in terms
of the covariant derivative:

\begin{itemize}
\item $c_1(f,g)= \left|\begin{array}{cc}\nabla f & \nabla g \\
\nabla^2 f & \nabla^2 g\end{array}\right|dz$ \\
\item $\bar{c}_1(f,g)= \left|\begin{array}{cc}f & g \\ \nabla^2 f &
\nabla^2 g\end{array}\right|dz^0$ \\
\item $c_2(f,g)=\left|\begin{array}{cc}\nabla f & \nabla g \\ \nabla^3
f &\nabla^3 g \end{array}\right|dz^2$ \\
\item $\bar{c}_2(f,g)=\left|\begin{array}{cc}f & g \\ \nabla^3 f
&\nabla^3 g\end{array}\right|dz^1$ \\ 
\item ${c}_5(f,g)=\left|\begin{array}{cc}\nabla^3 f & \nabla^3 g \\
\nabla^4 f & \nabla^4 g\end{array}\right|dz^5$
\item ${c}_7(f,g):=\left(2\left|\begin{array}{cc}\nabla^3 f & \nabla^3 g \\
\nabla^4 f & \nabla^4 g\end{array}\right|\,-\,9\left|\begin{array}{cc}\nabla^4 f & \nabla^4 g \\
\nabla^5 f & \nabla^5 g\end{array}\right|\right)dz^7$
\end{itemize}

The cocycle $c_2$ is the covariant derivative version of the
Krichever-Novikov cocycle; we learnt this expression from D. Millionshshikov.

Obviously, this description is much simpler. To show at least in
priciple how the proof of this coincidence looks like, take for
example $c_2$. We have to calculate

\begin{eqnarray*}
\nabla^3f &=& \nabla^2(f'\,+\,\Gamma f)\\
&=&\nabla(f''+\Gamma'f+\Gamma f') \\
&=& f'''+\Gamma''f + 2\Gamma'f' - \Gamma\Gamma'f-\Gamma^2f.
\end{eqnarray*}

Note that in the first line, $f$ is a vector field, but
$(f'\,+\,\Gamma f)$ is a function, in the second line,
$(f''+\Gamma'f+\Gamma f')$ is a 1-form and the result is a 2-form.

This gives

\begin{displaymath}
f(\nabla^3g)-g(\nabla^3f)\,=\,fg'''-gf'''\,+\,(2\Gamma-\Gamma^2)(fg'-gf')
\end{displaymath}

Identifying $(2\Gamma-\Gamma^2)$ with $(2\Gamma-\Gamma^2)=2R$, we get the
coincidence of the covariant derivative expression with $c_2$.

As said before, it is clear that all covariant derivative expressions
will be cocycles with values in the appropriate ${\cal F}_{\lambda}$ -
the computations are straight forward.

\section{Non-triviality of the cocycles}

Let us scetch here an argument showing the non-triviality of the
constructed cocycles:

Choose an embedding $S^1\to\Sigma_r$. The associated restriction of holomorphic
vector fields (resp. holomorphic $\lambda$-densities) to $S^1$ gives a
map $\phi: Hol(\Sigma_r)\to S^1$ (resp. $\psi:{\cal
F}_{\lambda}(\Sigma_r)\to {\cal F}_{\lambda}(S^1)$). These maps are
injective Lie algebra homomorphisms with dense image (where the image
is equipped with the induced topology from $Vect(S^1)$), cf
\cite{KriNov}.

There is a commutative diagram:\\

\setlength{\unitlength}{1cm}
\begin{picture}(12,3)
\put(0,0){$0$} \put(.5,.1){\vector(1,0){1}}
\put(1.7,0){${\cal F}_{\lambda}(S^1)$} \put(3,.1){\vector(1,0){1}}
\put(4.2,0){$Vect(S^1)\times{\cal F}_{\lambda}(S^1)$}
\put(7.6,.1){\vector(1,0){1}}
\put(9,0){$Vect(S^1)$}\put(10.4,.1){\vector(1,0){1}}\put(11.6,0){$0$}
\put(0,2){$0$} \put(.5,2.1){\vector(1,0){1}}
\put(1.7,2){${\cal F}_{\lambda}(\Sigma_r)$} \put(3,2.1){\vector(1,0){1}}
\put(4.2,2){$Hol(\Sigma_r)\times{\cal F}_{\lambda}(\Sigma_r)$}
\put(7.6,2.1){\vector(1,0){1}}
\put(9,2){$Hol(\Sigma_r)$}\put(10.4,2.1){\vector(1,0){1}}\put(11.6,2){$0$}
\put(2.2,1.7){\vector(0,-1){1}}\put(5.8,1.7){\vector(0,-1){1}}
\put(9.6,1.7){\vector(0,-1){1}}\put(9.2,1){$\phi$}\put(2.5,1){$\psi$}
\end{picture}\\

The non-triviality now follows from the non-triviality (see
\cite{OvsRog}) of the
corresponding cocycles for $Vect(S^1)$.

\end{document}